\documentclass[twocolumn,amsmath,amssymb,prl,groupedaddress,nobibnotes,showpacs]{revtex4-1}

\usepackage{graphicx}

\begin{document}

\title{Weak value measurement with an incoherent measuring device}

\author{Young-Wook Cho}
\email{soullio@postech.ac.kr}
\affiliation{Department of Physics, Pohang University of Science and Technology (POSTECH), Pohang, 790-784, Korea}

\author{Hyang-Tag Lim}
\affiliation{Department of Physics, Pohang University of Science and Technology (POSTECH), Pohang, 790-784, Korea}

\author{Young-Sik Ra}
\affiliation{Department of Physics, Pohang University of Science and Technology (POSTECH), Pohang, 790-784, Korea}

\author{Yoon-Ho Kim}
\email{yoonho@postech.ac.kr}
\affiliation{Department of Physics, Pohang University of Science and Technology (POSTECH), Pohang, 790-784, Korea}

\date{\today}

\begin{abstract}
In the Aharonov-Albert-Vaidman (AAV) weak measurement, it is assumed that the measuring device or the pointer is in a quantum mechanical pure state. In reality, however, it is often not the case. In this paper, we generalize the AAV weak measurement scheme to include more generalized situations in which the measuring device is in a mixed state. We also report an optical implementation of the weak value measurement in which the incoherent pointer is realized with the pseudo-thermal light. The theoretical and experimental results show that  the measuring device under the influence of partial decoherence could still be used for amplified detection of minute physical changes and are applicable for implementing the weak value measurement for massive particles.
\end{abstract}


\maketitle







\section{Introduction}
The projection postulate of the quantum theory states that the outcome of a measurement on a quantum system must be one of the eigenvalues of the measurement operator. The weak value introduced by Aharonov, Albert, and Vaidman, however, is quite peculiar in that the measurement outcomes of the weak value may lie well outside the normal range of the eigenvalues of the measurement operator \cite{AAV}. The weak value measurement, nevertheless, does not violate standard quantum theory and the effect is understood to be due to quantum interference of complex amplitudes \cite{Duck}.

The Aharonov-Albert-Vaidman (AAV) weak value measurement is accomplished in two steps: the weak measurement followed by postselection. The postselection step is the standard projection measurement (i.e., strong measurement) but, for the weak measurement, the measuring device or the pointer is assumed to be in a quantum mechanical pure state \cite{AAV,Duck}. In the case of quantum mechanical particles with mass whose center-of-mass coordinates are considered as the pointer for the measuring device, it becomes extremely difficult to achieve the measuring device in a pure state because the coupling to the environment causes decoherence of the pointer state as demonstrated, for example, in decoherence of matter waves \cite{Kleckner,Arndt,decoherence} and degradation of an atom laser beam \cite{atomlaser}. It is thus not surprising that the AAV weak value measurement to date has been implemented only with light whose spatial or temporal coherence can be used to represent the pointer in a pure state \cite{Ritchie,Resch,Solli,Wang,Hosten,lundeen,Dixon}.

In this paper, we generalize the AAV weak value measurement to include more generalized situations in which the measuring device (or the pointer state) is in a mixed state. We also report  an optical implementation of the weak value measurement in which the incoherent pointer is realized with the pseudo-thermal light. The theoretical and experimental results suggest that the measuring device under the influence of partial decoherence, i.e., the pointer state with partial coherence (or a density matrix with non-zero off-diagonal elements), could still be used for amplified detection of weak effects.

\section{Theory}

\subsection{Weak value measurement with a coherent measuring device}

We start with a brief description of the AAV weak value measurement \cite{AAV,Duck}. The impulse interaction Hamiltonian between the pointer and the system, whose observable $\hat A$ is to be measured, is given in general as 
\begin{equation}
\hat H = \delta ( t - t_0 )\hat p\hat A, \label{hamiltonian}
\end{equation}
where  $\hat p$ represents the momentum operator for the measuring device (with  the conjugate position operator $\hat q$) and $t_0$ is the time of measurement (i.e., interaction). In the AAV weak value measurement \cite{AAV,Duck}, the system is prepared in a pure state $|\psi_{in}\rangle$. The initial state of the pointer (i.e., the measuring device) is also assumed to be in a pure state (q-representation) as
\begin{equation}
\left| {\phi _{{{in}}} } \right\rangle  = \left(\frac{2}{\pi w_0^2}\right)^{1/4} \int {dq} \, \exp (-{q^2}/{w_0 ^2} ) | q \rangle, \label{md}
\end{equation}
where $w_0$ quantifies the pointer spread \cite{notep}.

After the interaction in Eq.~(\ref{hamiltonian}), the quantum state of both the system and the pointer is evolved to 
\begin{equation}
\exp{(-i \hat{p} \hat{A} / \hbar)} |\psi_{in}\rangle |\phi_{in}\rangle. 
\end{equation}
If we now make a projection measurement on the system in the $|\psi_f\rangle$ basis (i.e., postselection of the system having the quantum state $|\psi_f\rangle$), the pointer state is found to be \cite{notep}
\begin{eqnarray}
 \lefteqn{\langle {\psi _f }|  \exp{(-i \hat{p} \hat{A} / \hbar)} |\psi_{in}\rangle |\phi_{in}\rangle} \nonumber \\
  &\simeq& \left( {\langle \psi _f  | \psi _{in} \rangle  - i \hat p \langle {\psi _f } |\hat A| {\psi _{in} } \rangle/\hbar  +  \ldots } \right)| {\phi _{in} } \rangle  \nonumber\\
  &\simeq& N \langle {{\psi _f }} | {{\psi _{in} }} \rangle  \int{dp} \, \exp \left( - \frac{{w_0 ^2 p^2 + 4 i A_w p}}{{4 \hbar^2}} \right) | {p } \rangle,\label{evolve}
\end{eqnarray}
where $N \equiv  ({w_0^2}/{2\pi\hbar^2})^{1/4}$ and $A_w$ is the weak value defined as \cite{AAV,Duck}
\begin{equation}
A_w  \equiv \frac{\langle {\psi _f } | \hat A | {\psi _{in} } \rangle } {\langle {{\psi _f }} | {{\psi _{in} }} \rangle }. \label{wv}
\end{equation}
Note that  Eq.~(\ref{evolve}) has been derived with the assumption
\begin{equation}
\max_{n=2,3..} \left| \frac{{\langle \psi_f | \hat{A}^n | \psi_{in}  \rangle (p/\hbar)^n}}{{\langle \psi_f |\psi_{in}  \rangle}} \right| \ll |p A_w /\hbar| \ll 1.\label{wvlimit}
\end{equation}

Using Eq.~(\ref{md}), we can re-write Eq.~(\ref{evolve})  as 
\begin{equation}
\left(\frac{2}{\pi w_0^2}\right)^{1/4} \langle {{\psi _f }} | {{\psi _{in} }} \rangle \int {dq} \, \exp\left[ - \frac{(q - A_w )^2}{ w_0 ^2} \right] | q \rangle.\label{wvm}
\end{equation}
The essence of the weak value measurement is illustrated in Eq.~(\ref{wvm}): the pointer displays, as an outcome of the measurement, the weak value $A_w$ which may be much larger than any eigenvalues of $\hat A$ if $| {\psi _{in} } \rangle$ and $| {\psi _{f} } \rangle$ are nearly orthogonal to each other.

Although approximations were used to derive Eq.~(\ref{wvm}), it is in fact possible to calculate the effect of the AAV weak value measurement without any approximation. By expanding $|\psi_{in}\rangle$ and $|\psi_f\rangle$ in the eigenbasis of $\hat A$ as $| {\psi _{in} } \rangle  = \sum\nolimits_k {\alpha _k | {a_k } \rangle}$ and $| {\psi _f } \rangle = \sum\nolimits_{l} {\beta_{l} | {a_{l} } \rangle}$, the probability distribution $P_{\psi}(q)$ of the pointer $q$ is explicitly calculated to be
\begin{eqnarray}
P_{\psi}( q ) &=& \left| {\langle q |\langle {\psi _f } | \hat U | {\psi _{{in}} } \rangle | {\phi _{{in}} } \rangle } \right|^2  \nonumber \\
&=&\left| \left(\frac{2}{\pi w_0^2}\right)^{1/4}  \sum\limits_{k} {\alpha_k} {\beta_k^{*}} \exp \left[ - \frac{(q-a_k )^2}{w_0^2} \right] \right| ^{2}   \nonumber\\
&=&  \sqrt{\frac{2}{\pi w_0^2}} \sum\limits_{k,j} {\alpha _k } \beta_k^{*} \alpha_j^* \beta_j  \nonumber \\
& & \times \exp \left[ - \frac{(q - a_k )^2 +(q - a_j )^2 }{ w_0 ^2 }\right], \label{purecase}
\end{eqnarray}
where $\hat U=\exp{(-i \hat{p} \hat{A} / \hbar)}$, $a_j$ and $a_k$ are the eigenvalues of $\hat A$, and   we have used the orthonormality condition ${\langle {a _l} | {{a _k }} \rangle }=\delta _{l,k}$. Note that, in the weak value measurement limit shown in Eq.~(\ref{wvlimit}), $P_\psi(q)$ in Eq.~(\ref{purecase}) approximates to a single Gaussian peaked at the weak value $A_w$.

\subsection{Weak value measurement with an incoherent measuring device}

So far, we have considered the case in which the measuring device is in a pure state, i.e., the pointer spread is completely coherent as shown in Eq.~(\ref{md}). Let us now generalize the problem by considering that the measuring device (having the same pointer spread $w_0$) is no longer in a pure state, rather in a mixed state with some partial coherence quantified with $w_c$. The pointer state is then expressed as a density matrix 
\begin{eqnarray}
\rho _\phi   &=& \frac{\sqrt{2}}{\pi w_0 w_c}\int dq_0 dq'dq'' \exp \left[ - \frac{{{q_0 ^2 } }} {{w_0 ^2 }} \right] \nonumber \\
& \times &  \exp \left[ - \frac{{{(q' - q_0 )^2 } }} {{w_c ^2 }} \right]  \exp \left[ - \frac{{{(q'' - q_0 )^2 } }} {{w_c ^2 }} \right] | {q'} \rangle \langle {q''} | ,
\end{eqnarray}
and the initial system-pointer quantum state is described as 
\begin{equation}
| {\psi _{in} } \rangle \rho _\phi \langle {\psi _{in} } |.
\end{equation}

After the weak measurement, the initial system-pointer density matrix  $| {\psi _{in} } \rangle \rho _\phi \langle {\psi _{in} } | $ is evolved due to the interaction Hamiltonian in Eq.~(\ref{hamiltonian}) into 
\begin{equation}
\hat U | {\psi _{in} } \rangle \rho _\phi \langle {\psi _{in} } | \hat U^\dag. 
\end{equation}
Making a projection measurement on the system in  the $ | {\psi _f } \rangle $ basis (i.e., postselecting the system having the state $ | {\psi _f } \rangle $), the pointer state is found to be
\begin{eqnarray}
\rho_f &=& \langle {\psi _f } |\hat U| {\psi _{in} }
\rangle \rho _\phi  \langle {\psi _{in} } |\hat U^\dag  | {\psi _f } \rangle \nonumber \\
&=&\sum\limits_{k,j,l,m}  \beta_l^{*} \alpha _k \alpha _{j}^{*} }{\beta_{m} \langle {a _l } |\hat U| {a _{k} } 
\rangle \rho _\phi  \langle {a _{j} } |\hat U^\dag  | {a _m } \rangle.
\end{eqnarray}
The probability distribution for the pointer $P_\rho(q)$ is then calculated as
\begin{eqnarray}
P_\rho(q) &=& \langle q | \rho_f | q\rangle  \nonumber\\
&=& \frac{\sqrt{2}}{\pi w_0 w_c } \sum\limits_{k,j} \alpha _k \beta_k^{*} \alpha_j^{*} \beta_{j} \int dq_0 \nonumber \\
& & \times \exp\left[-\frac{q_0^2}{w_0^2} -\frac{(q-a_k-q_0)^2}{w_c^2} -\frac{(q-a_j-q_0)^2}{w_c^2}\right]    \nonumber \\
&=& \sqrt{\frac{2}{\pi (2w_0^2+w_c^2)}} \sum\limits_{k,j} \alpha _k \beta_k^{*} \alpha_j^{*} \beta_{j} \nonumber \\
& & \times \exp \left[ w_0 ^{-2} \left(  - \frac{( q - a_k )^2 } {\gamma^2 }  \right.\right.    - \frac{( q - a_j )^2 } {\gamma^2}\nonumber \\
& & \left.\left. + \frac{(2q - a_k  - a_j)^2 } {\gamma^4  + 2\gamma^2} \right)\right], \label{mixedcase}
\end{eqnarray}

where we have used the orthonormality conditions ${\langle {a _l} | {{a _k }} \rangle }=\delta _{l,k}$ and ${\langle {a _j} | {{a _m }} \rangle }=\delta _{j,m}$. Note that the degree of partial coherence is defined as $\gamma \equiv w_c/w_0$.

Equation (\ref{mixedcase}) shows that the weak value effect should still be observable even though the measuring device (i.e., the pointer) is in a mixed state whose degree of partial coherence is quantified with $\gamma$. Note that the pointer is effectively in a pure state in the limit  $\gamma \gg 1$ and, in this limit, Eq.~(\ref{mixedcase})  approximates to Eq.~(\ref{purecase}).

\section{Experiment}

\begin{figure}[tp]
\centering
\includegraphics[width=3.2in]{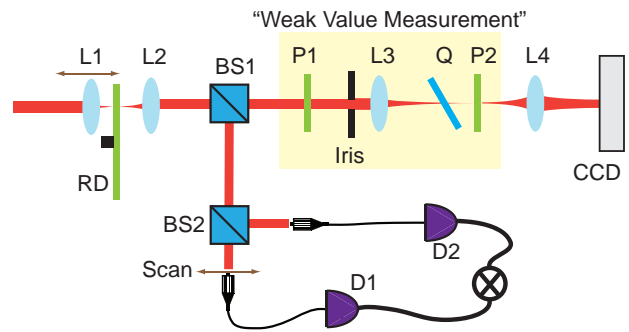}
\caption{Schematic of the experiment. The incoherent pointer state is realized with a pseudo-thermal light source whose transverse spatial coherence can be varied. A He-Ne laser beam is focused on a rotating ground disk (RD) with a movable lens L1 and the scattered light is collimated with another lens L2. BS1 and BS2 are 50/50 beam splitters. The transverse spatial coherence of the beam  is measured with detectors D1 and D2 and is used for determining $\gamma$, the degree of partial coherence of the pointer state. P1 and P2 are polarizers for state preparation and postselection, respectively. The weak measurement occurs at the tilted quartz plate Q.}\label{setup}
\end{figure}

The weak value measurement setup which incorporates the  pointer in a mixed state is schematically shown in Fig.~\ref{setup}. The system state is the polarization state of the photon (analogous to a spin 1/2 particle) and is assumed to be in a pure state. The transverse position of the photon corresponds to the pointer (i.e., the measuring device) for measuring the system state \cite{Duck,Ritchie}. 

The incoherent pointer state is realized with the pseudo-thermal light source based on scattering of a focused laser beam (a He-Ne laser operating at 632.8 nm) at a rotating ground disk (RD) \cite{Martienssen}. The focusing (L1) and collimating (L2) lenses have 30 mm and 75 mm focal lengths, respectively. By moving L1 longitudinally, thereby changing the beam size on the rotating ground disk, the degree of transverse spatial coherence of the collimated beam can be varied. Therefore, we can easily adjust the degree of partial coherence $\gamma$ of the pointer state in Eq.~(\ref{mixedcase}) by simply moving the focusing lens L1. The collimated beam is then split into two  by a beam splitter (BS1); one beam is for the weak value measurement and the other is for characterizing the pointer state.

For the weak value measurement, the photon is prepared in a definite polarization state with polarizer P1. An iris placed after P1 defines the $e^{-2}$ beam waist radius $w'_0=0.697$ mm. The lens L3 ($f= 100$ mm) then focuses the beam so that the beam waist is $w_0=\lambda f/(\pi w'_0)=28.9$ $\mu$m at the focus. The weak measurement on the system (i.e., the polarization state of the photon) is then implemented with a 0.5 mm thick quartz plate Q with its optic axis oriented vertically. The quartz plate Q causes a small polarization dependent displacement between the two orthogonal polarization components due to the birefringence. The expected displacement is much smaller (in our case $a=1.316$ $\mu$m) than the beam width $w_0$ and, therefore, the quartz plate Q acts as the weak measurement operator on the system \cite{Duck,Ritchie}.  In order to ensure that the incoming polarization of the photon is not changed by Q, i.e, the net phase difference of $2\pi$ between the vertical and horizontal polarization of the photon, Q  was tilted (about the optic axis) to $43.5^\circ$.

After the weak measurement by quartz plate Q, the postselection measurement (on polarization) is implemented by the polarizer P2 placed at the focus of L3. Finally, an imaging lens L4 ($f=50$ mm) and a CCD camera are used to measure the transverse spatial profile of the photon at the P2 location.

\section{Results and analysis}

\subsection{Characterizing the incoherent measuring device (pointer state)}

As mentioned earlier, we can vary the transverse spatial coherence of the collimated beam by changing the beam size on the rotating ground disk and the degree of transverse spatial coherence is directly related to the degree of partial coherence $\gamma$ in Eq.~(\ref{mixedcase}). Therefore, the first step in experimentally demonstrating the weak value measurement with an incoherent measuring device is to properly and accurately characterize the transverse spatial coherence of the collimated beam. 

In the experiment, the light scattered at the rotating ground disk RD is collimated with L2. Beam splitter BS2 then splits the collimated beam: the transmitted beam is used for the weak value measurement and the reflected beam is used for measuring the transverse spatial coherence of the pointer, see Fig.~\ref{setup}. The degree of transverse spatial coherence of the beam is measured with a Hanbury-Brown$-$Twiss type interferometer, consisting of a 50/50 beam splitter (BS2) and two detectors D1 and D2. Detectors D1 and D2 are multi-mode fiber coupled so that the effective diameter of the detectors is 62.5 $\mu$m, the core diameter of the fiber. The fiber connected to D1 can be scanned and the photocurrents from the detectors D1 and D2 are digitized and stored on a computer. The cross-correlation between the two split modes (of BS2) are then calculated using the AC components of the digitized photocurrents, $\Delta I (t)$, using the relation
\begin{equation}
C(x)=\frac{\langle \Delta I_1(x,t) \Delta I_2(t)\rangle_t }{\sqrt{\langle(\Delta I_1(x,t))^2\rangle_t} \sqrt{\langle(\Delta I_2(t))^2\rangle_t}},
\end{equation}
where the subscripts 1 and 2 refer to detectors D1 and D2, respectively, and $\langle \ldots \rangle_t$ represents time averaging. 

\begin{figure}[tp]
\centering
\includegraphics[width=3.3in]{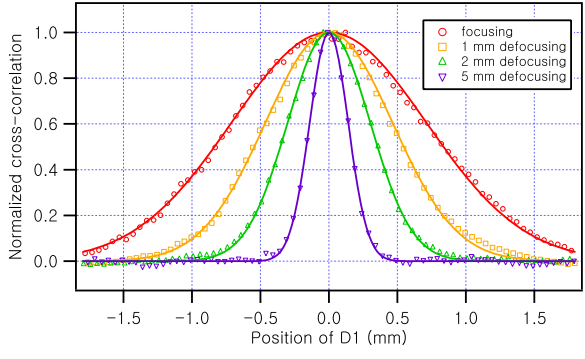}
\caption{The normalized cross-correlation, $C(x)$, as a function of D1 position, $x$, shows the degree of transverse spatial coherence of the collimated beam.  The measured $e^{-2}$ widths $w'_c$ are 1.42 mm, 0.93 mm, 0.60 mm, and 0.28 mm. The $w_c$ values for the weak value measurement can be calculated from $w'_c$ and they are 59.0 $\mu$m, 38.5 $\mu$m, 25.0 $\mu$m, and 11.7 $\mu$m, respectively. The degree of partial coherence $\gamma$ can then be calculated using the relation $\gamma = w_c/w_0$. See text for details.}\label{corr}
\end{figure}

The experimental results are shown in Fig.~\ref{corr}. The cross-correlation measurements show that defocusing of L1 causes reduction of the transverse spatial coherence of the collimated beam. The measured $e^{-2}$ widths $w'_c$ are 1.42 mm, 0.93 mm, 0.60 mm, and 0.28 mm, depending on the L1 position. Since the weak value measurement setup uses lens L3  ($f=100$ mm) to focus the beam, see Fig.~\ref{setup}, the measured value $w'_c$ should be converted to the value relevant in the weak value measurement setup using  the relation $w_c = w'_c  w_0 /w'_0$. As discussed in the previous section, $w'_0 = 0.698$ mm and $w_0 = \lambda f/(\pi w'_0)$ = 28.9 $\mu$m. The degree of partial coherence $\gamma$ in Eq.~(\ref{mixedcase}) is then calculated using the relation $\gamma = w_c/w_0$. The $\gamma$ values are  2.04 (focusing), 1.33 (1 mm defocusing), 0.865 (2 mm defocusing), and 0.404 (5 mm defocusing).

\subsection{Weak value measurement with an incoherent measuring device}

Since the degree of partial coherence $\gamma$ is determined for the measuring device (i.e., the pointer state),  we now can proceed to test weak value measurement with an incoherent measuring device. We start by re-writing the general result in Eq.~(\ref{mixedcase}) using experimentally relevant parameters.

In the experiment, the initial and final polarization states of the photon are assumed linear (P1 and P2 are linear polarizers) so that $| {\psi _{in} } \rangle  = \cos \alpha | H \rangle  + \sin \alpha | V \rangle$ and $ | {\psi _f } \rangle  = \cos \beta | H \rangle  + \sin \beta | V \rangle$. Since $|\psi_{in}\rangle$ and $|\psi_f\rangle$ should be almost orthogonal to observe the weak value effect, P1 and P2 angles are set at $\alpha=\pi/4$ and $\beta=-\pi/4+\epsilon$, respectively. Also, the eigenvalues of the observable, corresponding to the expected beam shift for each polarization state, are $a_H  =  - a$ and $a_V  = 0$. Under these conditions, Eq.~(\ref{mixedcase}) can be re-written as
\begin{eqnarray}
P_\rho(q) &\propto& \cos^2 \beta \exp\left[ w_0 ^{-2} \left( - \frac{2(a + q)^2} {\gamma^2} + \frac{4(a + q)^2}{\gamma^4  + 2\gamma^2} \right) \right] \nonumber\\ 
& &+ \sin 2\beta \exp \left[ w_0 ^{-2} \left( - \frac{a^2  + 2aq + 2q^2 } {\gamma^2} \right.\right. \nonumber \\ 
&&+ \left.\left.\frac{(a + 2q)^2 } {\gamma^4  + 2\gamma^2 } \right) \right] \nonumber \\
& & + \sin ^2 \beta \exp\left[ w_0 ^{-2} \left( { - \frac{{2q^2 }} {{\gamma^2 }} + \frac{{4q^2 }} {{\gamma^4  + 2\gamma^2 }}} \right) \right]. \label{theory}
\end{eqnarray}

%

\begin{figure}[tp]
\centering
\includegraphics[width=3.30in]{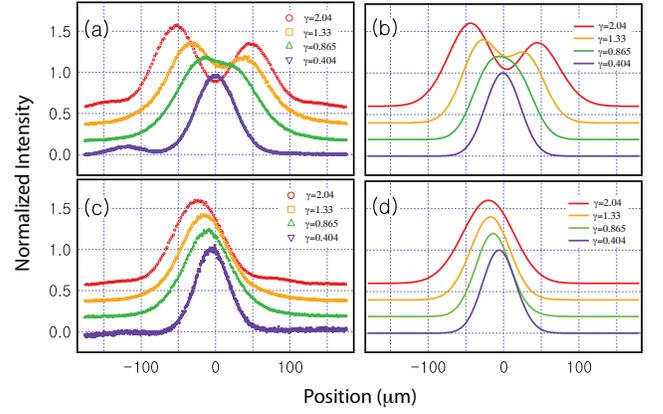}
\caption{Weak value measurement with an incoherent measuring device. (a) and (c) are experimental data and (b) and (d) are corresponding theoretical results plotted using  Eq.~(\ref{theory}). For (a) and (b), $\epsilon= 1.00 \times 10^{ - 3}$ rad. For  (c) and (d), $\epsilon= 2.79\times10^{ - 2}$ rad. All plots are normalized to unity and vertically shifted for clarity.
 }\label{weakvalue}
\end{figure}

We have performed the weak value measurement  with the incoherent pointer for several values of $\gamma$, which characterizes the  degree of partial coherence of the pointer (i.e., the measuring device), and for several values of $\epsilon$, which determines the $\langle \psi_f | \psi_{in}\rangle$ value. Note that, since the weak value $A_w$ does not make sense if $\langle \psi_f | \psi_{in}\rangle = 0$,  $\epsilon$ should not be zero. (If $\langle \psi_f | \psi_{in}\rangle = 0$, the weak value is not defined by the definition Eq.~(\ref{wv}), and the assumption Eq.~(\ref{wvlimit}) cannot be not satisfied.) To result a large weak value $A_w$, however, $\epsilon$ should be close to zero.

The experimental results and corresponding theoretical results are shown in Fig.~\ref{weakvalue}. In experiment, the peak position of the measured transverse spatial profile of the beam on the CCD represents the weak value $A_w$. The experimental results show that the weak value $A_w$, which is  larger than the eigenvalue of the operator $a_H=-a=-1.316$ $\mu$m, is observable even with the pointer (measuring device) in a mixed state if the pointer has some degree of partial coherence, i.e., nonzero off-diagonal elements of the density matrix representing the pointer state. Note also that, the larger the degree of partial coherence $\gamma$, the larger the resulting weak value $A_w$.

The experimental data also show that,  if $\epsilon$ is too close to zero ($\epsilon = 1.00 \times 10^{-3}$ rad), the weak value is not well defined, see Fig.~\ref{weakvalue}(a). The spatial profile shows two peaks when the $\gamma$ is large enough, but it reduces to a single Gaussian peak centered nearly at zero when  $\gamma$ is much smaller than 1. This clearly is  due to the lack of quantum interference.

The weak value effect is more clearly visible for a slightly larger value of $\epsilon$, $\epsilon  = 2.7 \times 10^{ - 2}$ rad. As shown in Fig.~\ref{weakvalue}(c), the weak value effect is reduced gradually as $\gamma$ gets smaller. Even for a rather small value $\gamma=0.404$ (an incoherent pointer with small partial coherence), a rather large  weak value $A_w=-4.58$ $\mu$m is observed. The experimental results, Figs.~\ref{weakvalue}(a) and \ref{weakvalue}(c), are in good agreement with the theoretical plots, Figs.~\ref{weakvalue}(b) and \ref{weakvalue}(d), calculated using Eq.~(\ref{theory}).

\begin{figure}[tp]
\centering
\includegraphics[width=3.3in]{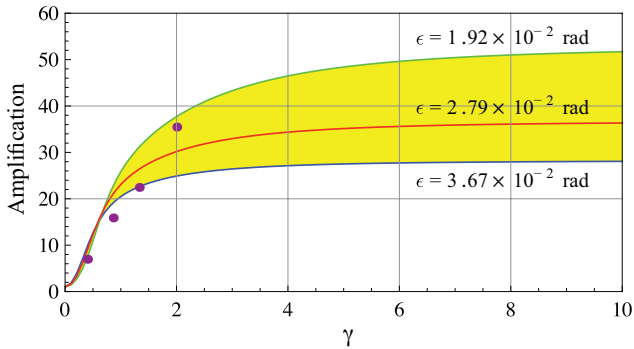}
\caption{Weak value amplification as a function of $\gamma$. Experimental data points are from Fig.~\ref{weakvalue}(c) for $\epsilon = 2.79 \times 10^{-2}$ rad. For each data point, the error bars are smaller than the solid circle. The solid lines are due to the theoretical result in Eq.~(\ref{theory}). The upper and lower solid lines are for the weak value amplification calculated for $\epsilon = 1.92 \times 10^{-2}$ rad and $\epsilon = 3.67 \times 10^{-2}$ rad, respectively. These $\epsilon$ values correspond to relative angle setting errors of  $\pm 0.5^\circ$ between polarizers P1 and P2 \cite{last}. }\label{amp}
\end{figure}

Finally, the weak value effect as a function of $\gamma$ is summarized in Fig.~\ref{amp}, which shows that the amplification (defined as the ratio between the peak position of the spatial profile and the  expectation value of the observable $\hat A$, i.e., $\langle \psi_{in} |\hat A|\psi_{in} \rangle = -a/2=-0.658$ $\mu$m) depends heavily on $\gamma$ and $\epsilon$. For $\gamma \gg 1$, the pointer state becomes effectively pure so that the amplification factor is bounded to a specific value. In the intermediate range of $\gamma$, the amplification factor increases as $\gamma$ gets larger. We would like to point out that the weak value amplification can still be observed even for $\gamma < 1$. Of course, if the pointer is completely incoherent (i.e., $\gamma=0$), no weak value effect can occur (i.e. no amplification).

\section{Conclusion} 
In conclusion, we have generalized the AAV weak value effect to include the situations in which the measuring device  (the pointer) is in a mixed state and have demonstrated the generalized weak value effect in an optical experiment in which the pointer in a mixed state is realized with the pseudo-thermal light source of a varying degree of partial spatial coherence. We have also introduced an experimentally measurable quantity which effectively quantifies the partial coherence of the pointer.  Our results show that the pointer state no longer in a pure state but in a mixed state (with some partial coherence) can still exhibit the weak value effect and thus may be used for amplified detection of a very small physical changes.  The result reported in this paper should be directly applicable to weak value measurement schemes involving a beam of massive particles whose pointer states (the transverse profile of the beam) cannot be expected to be in a pure state due to the decoherence causing interactions, such as, inter-particle collisions, strong coupling with the environment, etc.

\section*{Acknowledgements}
This work was supported by the National Research Foundation of Korea (2009-0070668) and POSTECH BSRI Research Fund. YWC, HTL, and YSR acknowledge partial support from the Korea Research Foundation (KRF-2008-314-C00075).


\begin{thebibliography}{99}
\bibitem{notep} In the $p$ representation, the pointer state is $|\phi_{in}\rangle  = ({w_0 ^2}/{2 \pi \hbar^2} )^{{1}/{4}} \int {dp} \, \exp (-{w_0 ^2 p^2}/{4\hbar^2} ) | p \rangle$ with the inner product $\langle p | q \rangle =  \exp (-ipq/\hbar)/\sqrt{2 \pi \hbar}$.
\bibitem{last} The angle setting error of $\pm 0.5^\circ$ comes from the fact that our rotation mounts were graded in $2^\circ$ increment.
\end{thebibliography}
\end{document}